\documentclass{article}
\makeatletter
\usepackage{makecell}
\usepackage[utf8]{inputenc}
\usepackage{xspace}
\usepackage{float}
\usepackage[caption = false]{subfig}
\usepackage{multirow}
\usepackage{graphicx}
\usepackage[margin=1in]{geometry}
\usepackage[T1]{fontenc}
\usepackage[bitstream-charter]{mathdesign}
\usepackage{amsmath}

\usepackage{scalefnt,letltxmacro}
\LetLtxMacro{\oldtextsc}{\textsc}
\usepackage[scaled=0.92]{PTSans}

\usepackage[usenames,dvipsnames]{xcolor}

\usepackage[english]{babel}
\usepackage[parfill]{parskip}
\usepackage{afterpage}
\usepackage{framed}

{\endMakeFramed}

\usepackage{ragged2e}

\newcounter{parcount}

\usepackage{graphicx}
\usepackage{fancyvrb}
\fvset{fontsize=\normalsize}

\usepackage[colorlinks,linktoc=all]{hyperref}
\usepackage[all]{hypcap}
\hypersetup{citecolor=Blue}
\hypersetup{linkcolor=Blue}
\hypersetup{urlcolor=Blue}

\usepackage[nameinlink]{cleveref}

\usepackage[acronym,smallcaps,nowarn]{glossaries}
\glsdisablehyper

\newacronym{PGD}{pgd}{projected gradient descent}
\newacronym{ECG}{ecg}{electrocardiogram}
\newacronym{EEG}{eeg}{electroencephalogram}
\newacronym{AF}{af}{atrial fibrillation}
\newacronym{FGSM}{fgsm}{fast gradient sign method}
\newacronym{SAP}{sap}{smooth adversarial perturbations}
\newacronym{PGHD}{pghd}{Patient-generated Health Data}
\newacronym{FDA}{fda}{Food and Drugs Administration}
\newacronym{RWD}{rwd}{Real-World Data}

\newcommand{\ECG}{\gls{ECG}\xspace}
\newcommand{\ECGs}{\glspl{ECG}\xspace}
\newcommand{\AF}{\gls{AF}\xspace}
\title{Adversarial Examples for Electrocardiograms}
\author{Xintian Han$^{1*}$, Yuxuan Hu$^2$, Luca Foschini$^3$, Larry Chinitz$^2$, Lior Jankelson$^2$, Rajesh Ranganath$^{1,4,5*}$\\
$^1$ Center for Data Science, New York University\\
$^2$ Leon. H. Charney Division of Cardiology, NYU Langone Health \\
$^3$ Evidation Health, Inc. \\
$^4$ Department of Population Health, NYU Langone Health \\
$^5$ Courant Institute of Mathematics, New York University
}
\date{}
\begin{document}
\maketitle
\textbf{In recent years, the \gls{ECG} has seen a large diffusion in both medical and commercial applications, fueled by the rise of single-lead versions. Single-lead \gls{ECG} can be embedded in medical devices and wearable products such as the injectable Medtronic Linq monitor, the iRhythm Ziopatch wearable monitor, and the Apple Watch Series 4. Recently, deep neural networks have been used to automatically analyze \gls{ECG} tracings, outperforming even physicians specialized in cardiac electrophysiology~\cite{rajpurkar2017cardiologist} in detecting certain rhythm irregularities. However, deep learning classifiers have been shown to be brittle to adversarial examples, which are examples created to look incontrovertibly belonging to a certain class to a human eye but contain subtle features that fool the classifier into misclassifying them into the wrong class~\cite{goodfellow2014explaining, szegedy2013intriguing}. Very recently, adversarial examples have also been created for medical-related tasks~\cite{finlayson2018adversarial,paschali2018generalizability}.   Yet, traditional attack methods to create adversarial examples, such as \gls{PGD} \cite{madry2017towards} do not extend directly to \gls{ECG} signals, as they generate examples that introduce square wave artifacts that are not physiologically plausible. Here, we developed a method to construct {\em smoothed} adversarial examples for single-lead \gls{ECG}. First, we implemented a neural network model achieving state-of-the-art performance on the data from the 2017 PhysioNet/Computing-in-Cardiology Challenge for arrhythmia detection from single lead \gls{ECG} classification~\cite{clifford2017af}. For this model, we utilized a new technique to generate smoothed examples to produce signals that are 1) indistinguishable to cardiologists from the original examples and 2) incorrectly classified by the neural network. Finally, we show that adversarial examples are not unique and provide a general technique to collate and perturb known adversarial examples to create new ones.} 

\textbf{Background}. Cardiovascular diseases represent a major health burden, accounting for 30\% of deaths worldwide~\cite{kelly2010promoting}. The \gls{ECG} is a simple and non-invasive test used for screening and diagnosis of cardiovascular disease. It is widely available in multiple medical device applications, including standard 12-lead \gls{ECG}, Holter recorders, and monitoring devices~\cite{kennedy2013evolution}. In recent years, there has been further growth in \gls{ECG} utilization in the form of
single-lead \gls{ECG} used in miniature implantable medical devices and wearable medical consumer products such as smart watches. These single-lead \glspl{ECG}, such as the one incorporated in the Apple Watch Series 4, are expected to be worn by tens of millions of Americans by the end of 2019~\cite{IDC2018}. Moreover, consumer wearable devices are utilized to collect data in clinical studies, such as the Health eHeart study~\cite{UCSF2018} and the Apple Heart Study~\cite{ACC2019}.
Large studies that make use of \gls{PGHD} are expected to become more frequent after the \gls{FDA}'s recent release of a set of guidelines and tools to collect \gls{RWD} from research participants via apps and other mobile health sources~\cite{FDARWE}. Having clinicians analyze such a large number of \glspl{ECG} is impractical.
Recently, driven by the introduction  of deep learning methodologies, automated systems have been developed, allowing rapid and accurate \gls{ECG} classification~\cite{hannun2019cardiologist}. In the 2017 PhysioNet Challenge for atrial fibrillation classification using single-lead \gls{ECG}, multiple  efficient solutions utilized deep neural networks~\cite{hong2017encase}. Deep learning has been shown to be susceptible to adversarial examples in general~\cite{goodfellow2014explaining,szegedy2013intriguing} and very recently in medical applications~\cite{finlayson2019adversarial}. However, to the best of our knowledge, it is unknown whether deep learning algorithms are robust in \ECG classification.

\paragraph{Description of data.}
    % Where does the data come from, single lead remove ECG, length of 
    % data in seconds on average. Counts of 4 classes of labels (Normal)
    % Probably don't need mean time-length 
    % Test, training size, validation 10%, 90%
\glspl{ECG} were obtained from the publicly available 2017 PhysioNet/CinC Challenge~\cite{clifford2017af}. The goal of the challenge was to classify single lead \ECG recordings to four types: normal sinus rhythm (Normal), \AF, an alternative rhythm (Other), or noise (Noise). The challenge data set contained 8,528 single-lead \ECG recordings lasting from 9s to about 60s, including 5,076 Normal, 758 \AF, 2,415 Other, and 279 Noise examples. 90\% of the data set was used for training and 10\% was used for testing.

\paragraph{Model and Performance.}
We used a 13-layer convolutional network~\cite{goodfellowtowards} that won the 2017
PhysioNet/CinC Challenge.
We evaluated both accuracy and F1 score. A high F1 score indicates good network performance, with high true positive and true negative rates. 

The model achieved an average accuracy rate of 0.88 and F1 score of 0.87 for the \ECG classes (Normal, \AF and Other) on the test set, which is comparable to state of the art \ECG classification systems~\cite{goodfellowtowards}. 

\begin{figure}[!ht]
    \centering
\subfloat[AF to Normal]{\includegraphics[width=\textwidth]{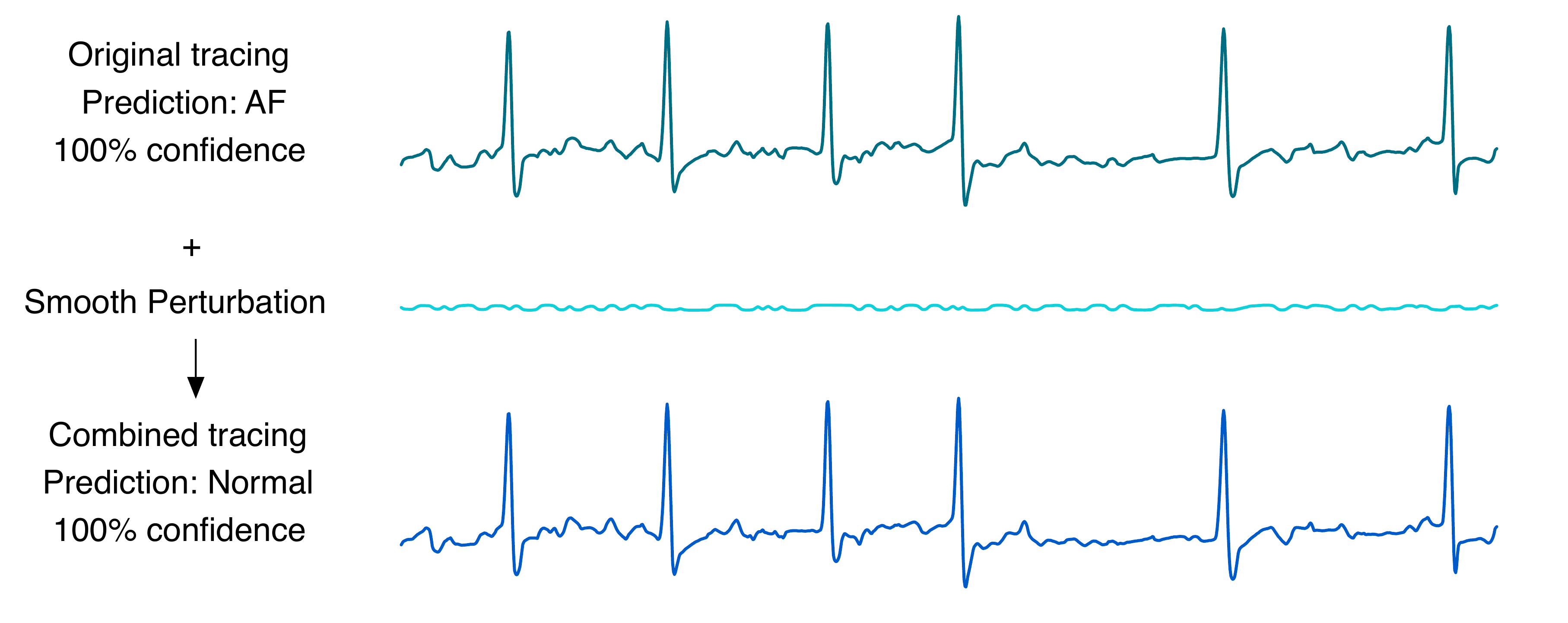}}\\
\subfloat[Normal to AF]{\includegraphics[width=\textwidth]{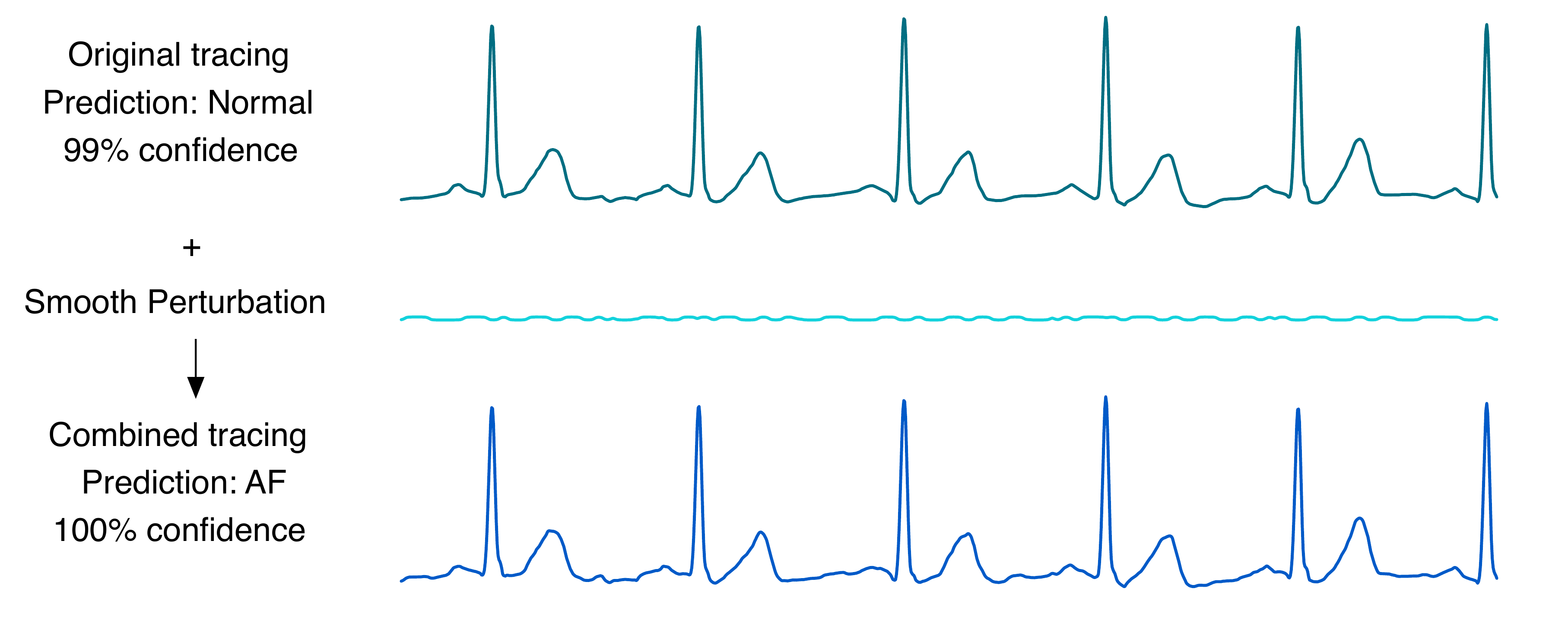}}
\caption{\label{fig:demo} Demonstration of disruptive adversarial examples. We add fine smooth perturbations to the original tracings to create adversarial examples. Perturbation and tracing voltage are plotted at the same scale. The neural network correctly diagnose normal sinus rhythm and \gls{AF} with 99\% and 100\% confidence when the original tracings are used. However, when presented with adversarial examples indistinguishable to expert clinician, the neural network misclassifies normal sinus rhythm to \gls{AF} and \gls{AF} to normal sinus with 100\% confidence.}
\end{figure}

\paragraph{Adversarial Examples.}
Adversarial examples are designed by humans to cause a machine learning algorithm to make a mistake. An adversarial example is made by adding a small perturbation to the input of the machine learning algorithm that keeps the label of the input, while also ensuring it still looks like a real input~\cite{goodfellow2014explaining,szegedy2013intriguing}. These kinds of adversarial examples have been successfully created in the field of medical imaging classification~\cite{finlayson2018adversarial}.

Traditional adversarial attack algorithms add a small imperceptible perturbation to lower the prediction accuracy of a machine learning model.

However, attacking \ECG deep learning classifiers with traditional methods creates examples that display square wave artifacts that are not physiologically plausible (Extended Data \cref{fig:pgd}).
By taking weighted average of nearby time steps, we crafted smooth adversarial examples that cannot be distinguished from original \ECG signals but will still fool the deep network to make a wrong prediction (See Methods section). 

\begin{figure}[!ht]
    \centering
    \includegraphics[width=.95\textwidth]{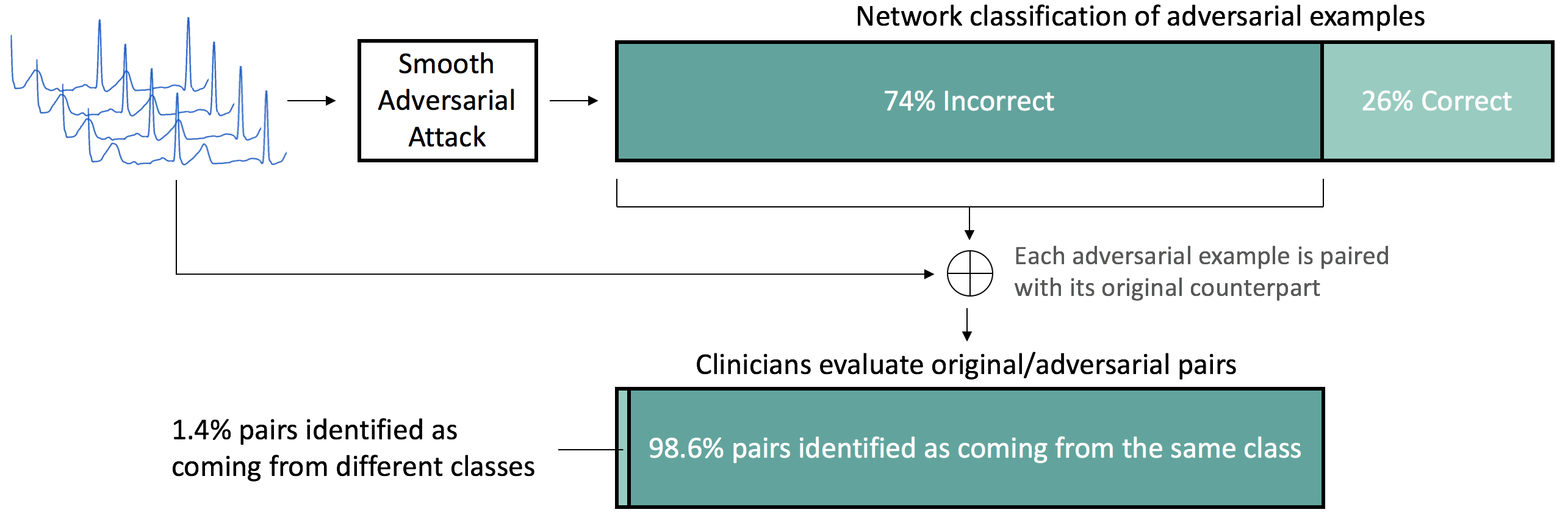}
    \caption{\label{fig:prog} Network accuracy on our generated adversarial examples. We are able to generate adversarial examples, leading to missclasification in 74\% of \gls{ECG} tracings. We validate the generated tracings by asking clinicians to determine whether 250 pairs of adversarial examples and original \ECGs come from the same class. On average, the clinicians conclude that \textbf{246.5/250} pairs belonged to the same class, confirming that the attacks did not substantially modify the tracings. This shows that the deep network failed to correctly classify most of the newly generated examples, when a human would have assigned to a different class only 1.4\% of them.} 
\end{figure}

We generated adversarial examples on the test set.
We transformed the test examples to make the network change the label of Normal, Other and Noise to any other label. For \AF, we altered the \AF test examples so that the deep neural network classifies them as Normal.
Misdiagnosis of \AF as Normal may increase the risk of \AF-related complications such as stroke and heart failure. We showcase the generation of adversarial examples in \Cref{fig:demo}.

\begin{figure}[!ht]
\centering
\includegraphics[width=.95\textwidth]{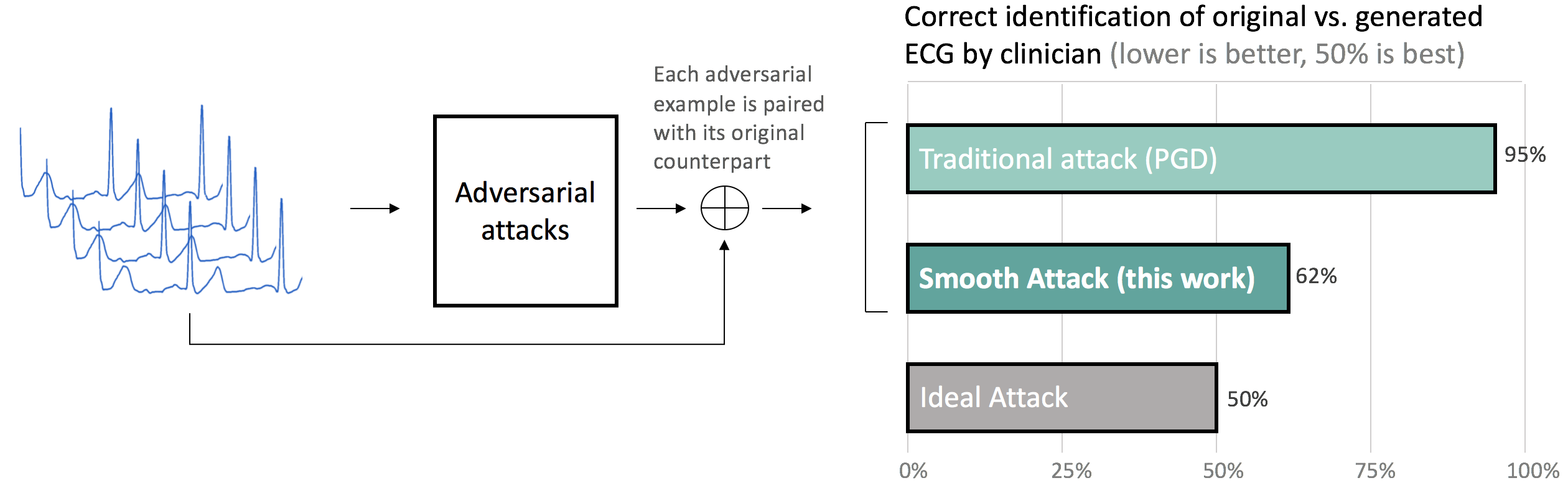}
\caption{\label{fig:acc}Clinician success rate of distinguishing real \ECG's from adversarial examples. We asked clinician experts in \ECG reading to distinguish between 100 pairs of original \ECGs and adversarial examples generated by the traditional attack method. They, on average,  correctly identified 95\% of the adversarial examples. We then asked them to distinguish between 100 pairs of original \ECGs and adversarial examples generated by our smooth attack method. This resulted in the correct identification of only 62\% of examples.}
\end{figure}

After adversarial attacks, 74\% of the test \glspl{ECG} originally classified correctly by the network are now assigned a different diagnosis, ultimately showing that deep \gls{ECG} classifiers are vulnerable to adversarial examples.

To assess how the generated signals would be classified by human experts, we invited one board certified medicine specialist and one cardiac electrophysiology specialist. We asked them to diagnose whether signals generated by our methods and original \glspl{ECG} come from the same class. From \Cref{fig:prog}, the model incorrectly diagnosed almost all (98.6\%) of the signals created by our method.

\begin{figure}[]
    \centering
    \includegraphics[width=.95\textwidth]{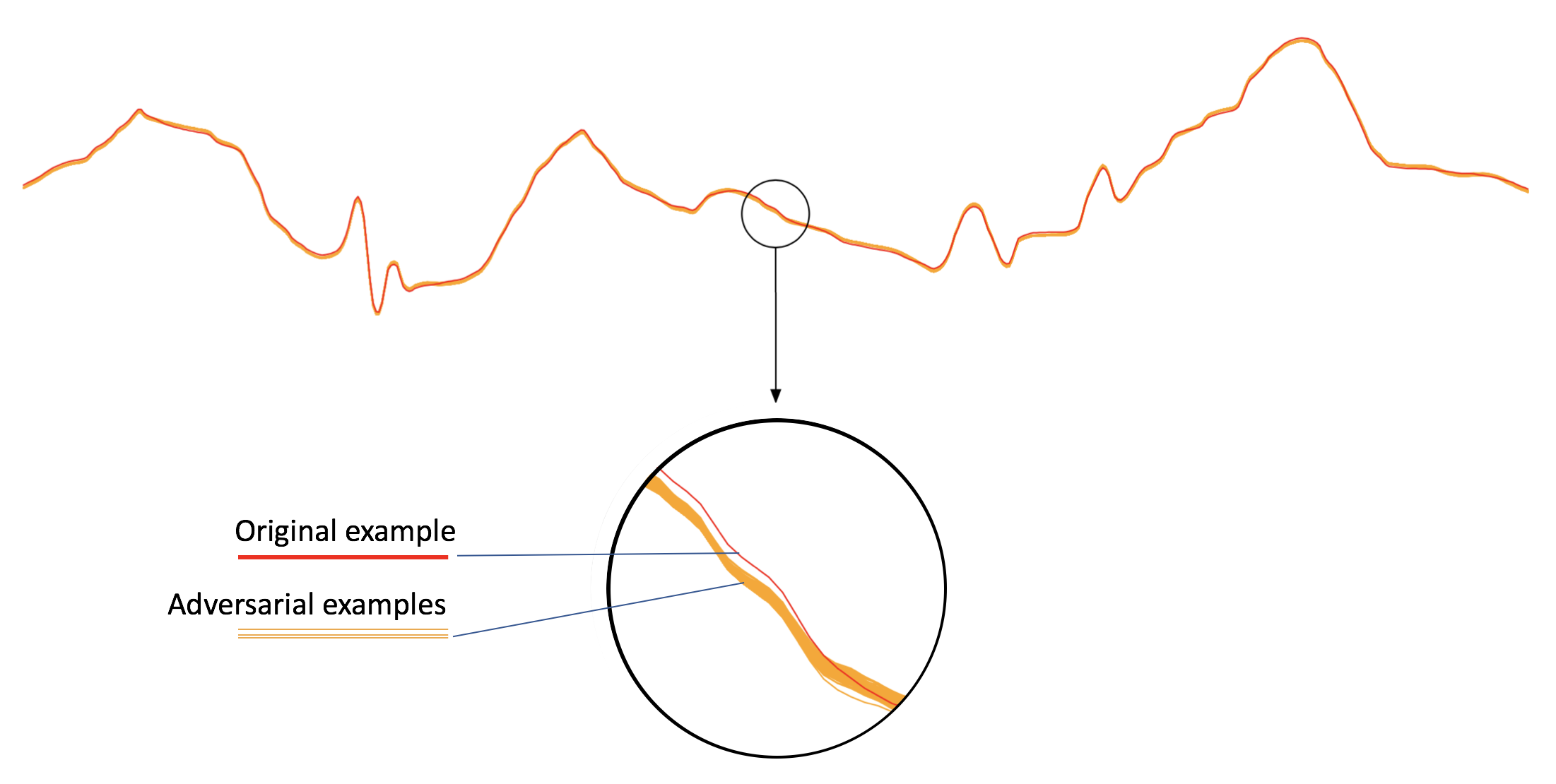}
        \caption{\label{fig:exist} Existence of adversarial examples. We picked an \ECG signal and we sampled 1,000 different adversarial examples by adding small Gaussian noise and smoothing. We plotted all the newly generated adversarial examples as well as the original \ECG signal. The newly generated examples form a wide band around the original signal. Sampling uniformly from the band and smoothing also produces adversarial examples.  These findings demonstrate that adversarial examples are not rare.}
\end{figure}

We also invited the clinical specialists to distinguish \ECG signals from the adversarial examples generated by our smooth method and the traditional attack method based on \glsreset{PGD}\gls{PGD}~\cite{madry2017towards}.
From \Cref{fig:acc}, the adversarial examples generated by our method are significantly harder for clinicians to distinguish from the original \gls{ECG} than the traditional attack method. On average, the clinicians were able to correctly identify the smoothed adversarial examples from their original counterpart 62\% of the time.\footnote{The EP specialist was slightly more accurate at 65\% versus 59\%.}

\paragraph{Existence of adversarial examples.}  
Here we provide a construction that shows
that adversarial examples are not rare. In particular, we show that it is possible to create more examples that remain adversarial by adding a small amount of Gaussian noise to an original adversarial example and then smooth the result. We repeat this 
process 1000 times and find that the deep neural network still incorrectly 
classifies all 1000 new, adversarial examples. Adding Gaussian noise could still produce adversarial examples on 87.6\% of the test examples from 
which adversarial examples were generated.
We plotted all of the newly crafted adversarial examples which form a band around the original \ECG signal in \Cref{fig:exist}.
The signals in the band may intersect.
We chose pairs of intersecting signals and concatenate the left half of one signal with right half of the other to create a new example. We found that signals created by concatenation are also adversarial examples. We also sampled random values in the band
for each time step and then smoothed them to create new adversarial examples. 
The fact that these different perturbations on adversarial examples all led to new examples that remain adversarial reinforces the notion that such examples should not be be considered rare isolated cases. (See Methods section for detailed description.)

\paragraph{Discussion.}
We demonstrated the ability to add imperceptible perturbations to \gls{ECG} tracings in order to create adversarial examples that fool a deep neural network classifier into assigning the examples to an incorrect rhythm class.
Moreover, we showed that such examples are not rare.
These findings raised several questions regarding the use
of deep learning in analyzing \glspl{ECG} at scale where millions of tests may be run every week by widespread consumer devices. To increase robustness to adversarial examples, it is crucial that classification methods for \glspl{ECG}, especially those intended to operate without humans supervision, generalize well to new examples. 
Ensuring safe generalization would require obtaining data acquired from multiple environments and from each new version of the signal collecting device. 
Labeling data for new environments and devices would also require substantial human effort, as clinical experts would be needed to provide correct labels for each new tracing.
The extent to which the rarity of examples increases as a function of training dataset size and composition warrants further investigation. 

One way to protect against adversarial examples is to include them in the training set of the model. However, such an approach can only protect against \emph{known} adversarial examples, created with a given specific attack method, and will not protect against future attack methods. 
A more direct approach would be to certify 
deep neural networks for robustness with mathematical proofs~\cite{singh2018fast} as suggested for other safety-critical domains, such as the aviation industry~\cite{julian2018deep}.

The possibility to construct even a single adversarial example may still enable malicious actors to inject small perturbations into real-world data indistinguishable to the human eye. This could represent an important vulnerability with implications including the attacks of medical devices relying on \ECG interpretation (pacemakers, defibrillators), the introduction of intentional bias into clinical trials, and the skew of data to alter insurance claims~\cite{finlayson2018adversarial}. To prevent such possibilities, it is paramount that platforms for collection and analysis of real-world data implement principles from Trusted Computing to provide trusted data provenance guarantees that can certify that data has not been tampered with from device acquisition to any downstream analysis~\cite{lyle2010trusted}.

One thing to note is that the lack of robustness observed is not inherent to the use of statistical methods to classify \glspl{ECG}. Humans tend to be more robust to small perturbations because they use coarser visual features to classify \glspl{ECG}, such as the R-R interval and the P-wave morphology. Coarser features change less under small perturbations and generalize better to new domains. 
These coarser features often have underlying biophysical meanings.
To automate the classification of less prevalent \gls{ECG} diagnoses, it may be useful to incorporate known electrocardiographic markers of biophysical phenomena along with deep learning to not only increase robustness to adversarial attacks but also improve the network accuracy. Additionally, regularizing deep networks to prefer coarser features can improve robustness.
In conclusion, with this work, we do not intend to cast a shadow on the utility of deep learning for \ECG analysis, which undoubtedly will be useful to handle the volumes of physiological signals available in the near future. This work should, instead, serve as an additional reminder that machine learning systems deployed in the wild should be designed with safety and reliability in mind~\cite{saria2019tutorial}, with particular focus on training data curation and provable guarantees on performances.

\section*{Acknowledgements}
We thank Wei-Nchih Lee, Sreyas Mohan, Mark Goldstein, Aodong Li, Aahlad Manas Puli, Harvineet Singh, Mukund Sudarshan and Will Whitney. 
\clearpage
\appendix
\section*{Methods}
\paragraph{\label{sec:attack}Description of the Traditional Attack Methods.}
Two traditional attack methods are \gls{FGSM}~\cite{goodfellow2014explaining} and \glsreset{PGD}\gls{PGD}~\cite{kurakin2016adversarial}. They are white-box attack methods based on the gradients of the loss used to train the model with respect to the input.

Denote our input entry $x$, true label $y$, classifier (network) $f$, and loss function $L(f(x),y)$. We describe \gls{FGSM} and \gls{PGD} below:
\begin{itemize}
\item \glsreset{FGSM}\Gls{FGSM}. \gls{FGSM} is a fast algorithm. For 
an attack level $\epsilon$, \gls{FGSM} sets
\[
x_{\text{adv}} = x+ \epsilon \cdot \text{sign}(\nabla_x L(f(x),y)),
\]
The attack level is chosen to be sufficiently small so as to be undetectable.
\item \glsreset{PGD}\Gls{PGD}. \gls{PGD} is an improved version that uses multiple iterations of \gls{FGSM}. Define $Clip_{x,\epsilon}(x')$ to project each $x'$ back to the infinite norm ball by clamping the maximum absolute difference value between $x$ and $x'$ to $\epsilon$. Beginning by setting $x_0' = x$, we have 
\begin{equation}
\label{eqn:x}
x_i' = Clip_{x,\epsilon}\{x_{i-1}'+\alpha \cdot\text{sign}(\nabla_x L(f(x_{i-1}'),y))). \}	
\end{equation}
After $T$ steps, we get our adversarial example $x_{\text{adv}} = x_T'$.  
\end{itemize}
\paragraph{\label{sec:smooth}Our Smooth Attack Method.}

In order to smooth the signal, we use the help of convolution. By convolution, we take the weighted average of one position of the signal and its neighbors:
\[
(a \circledast v)[n] = \sum_{m = 1}^{2K+1} a[n-m+K+1]\cdot v[m],
\]
where $a$ is the objective function and $v$ is the weights or kernel function. In our experiment, the weights are determined by a Gaussian kernel. Mathematically, if we have a Gaussian kernel of size 2K+1 and standard deviation $\sigma$, we have 
\begin{align*}
v[m] = \frac{\exp(-(m-K-1)^2/(2*\sigma^2))}{\sum_{i=1}^{2K+1}\exp(-(i-K-1)^2/(2*\sigma^2))}.
\end{align*}
 We can easily see that when $\sigma$ goes to infinity, the convolution with Gaussian kernel becomes a simple average; when $\sigma$ goes to zero, the convolution becomes an identity function.
Instead of getting an adversarial perturbation and then convolving it with the Gaussian kernels, we could create adversarial examples by optimizing a smooth perturbation that
fools the neural network.
We introduce our method of training \gls{SAP}. In our \gls{SAP} method, we take the adversarial perturbation as the parameter $\theta$ and add it to the clean examples after convolving with a number of Gaussian kernels. We denote $K(s,\sigma)$ to be a Gaussian kernel with size $s$ and standard deviation $\sigma$. The resulting adversarial example could be written as a function of $\theta$:
\[
x_{\text{adv}}(\theta) = x + \frac{1}{m}\sum_i^m\theta \circledast K(s[i], \sigma[i]) .
\]
In our experiment, we let $s$ be $\{5,7,11,15,19\}$ and $\sigma$ be $\{1.0, 3.0, 5.0, 7.0, 10.0\}$.
Then we try to maximize the loss function with respect to $\theta$ to get the adversarial example. We still use PGD but on $\theta$ this time:
\begin{equation}
\label{eqn:theta}
\theta_i' = Clip_{0,\varepsilon}\{\theta_{i-1}'+\alpha \cdot\text{sign}(\nabla_{\theta} L(f(x_{\text{adv}}(\theta_{i-1}'),y))) \}.
\end{equation}
There are two major differences between updates \eqref{eqn:theta} and \eqref{eqn:x}. In \eqref{eqn:theta}, we update $\theta$ not $x_{\text{adv}}$ and clip around zero not the input $x$. In practice, we initialize the adversarial perturbation $\theta$ to be the one obtained from \gls{PGD} ($\epsilon = 10, \alpha = 1, T = 20$) on $x$ and run another \gls{PGD} ($\epsilon = 10, \alpha = 1, T = 40$) on $\theta$. 

\paragraph{Existence of Adversarial Examples}
We design experiments to show that adversarial examples are not rare. Denote original signal to be $x$ and adversarial example we generated to be $x_{\text{adv}}$. 

First, we generate Gaussian noise $\delta \sim \mathcal{N}(0,25)$ and then add it to the adversarial examples. To make sure the new examples are still smooth, we smooth the perturbation by convolving with the same Gaussian kernels in our smooth attack method. We then clip the perturbation to make sure that it is still in the infinite norm ball. The newly generated example is 
\[
x_{\text{adv}}'= x + Clip_{0,\epsilon}\left\{\frac{1}{m}\sum_{i=1}^m (x_{\text{adv}}+\delta-x)\circledast K(s[i], \sigma[i])\right\}.
\]
We repeat the process of generating new examples 1000 times. These newly generated examples are still adversarial examples. Some of them may intersect. For each intersected pair, we concatenate the left part of one examples and the right part of the other to create new adversarial examples. Denote $x_1$ and $x_2$ to be a pair of adversarial examples that intersect. Suppose they intersect at time step $t$ and the total length of the example is $T$. The new hybrid example $x'$ satisfies:
\[
x'[1:t] = x_1[1:t]; \quad x'[t+1:T] = x_2[t+1:T],
\]
where $[1:t]$ means from time step $1$ to time step $t$. All the newly concatenated examples are still misclassified the network.

\begin{figure}[!ht]
    \centering
\subfloat[Adding Gaussian noise]{\includegraphics[width=\textwidth]{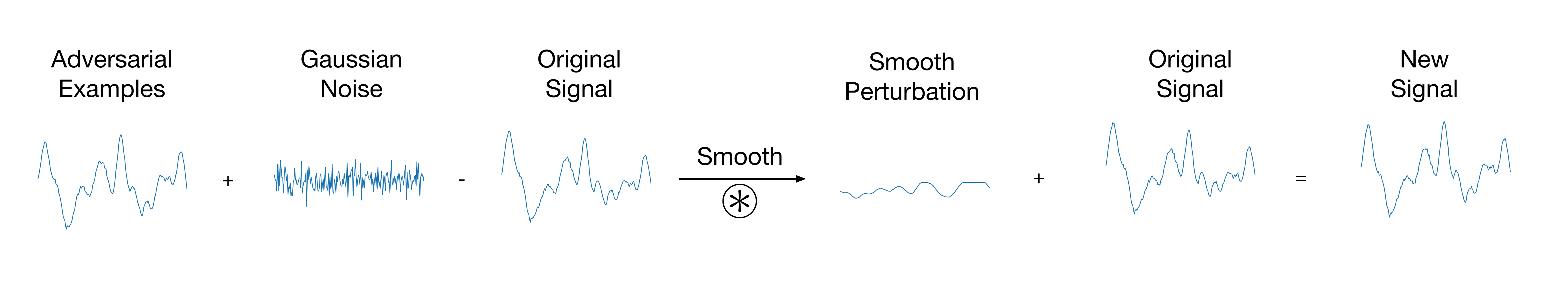}}\\
\subfloat[Concatenating intersected pairs]{\includegraphics[width=0.5\textwidth]{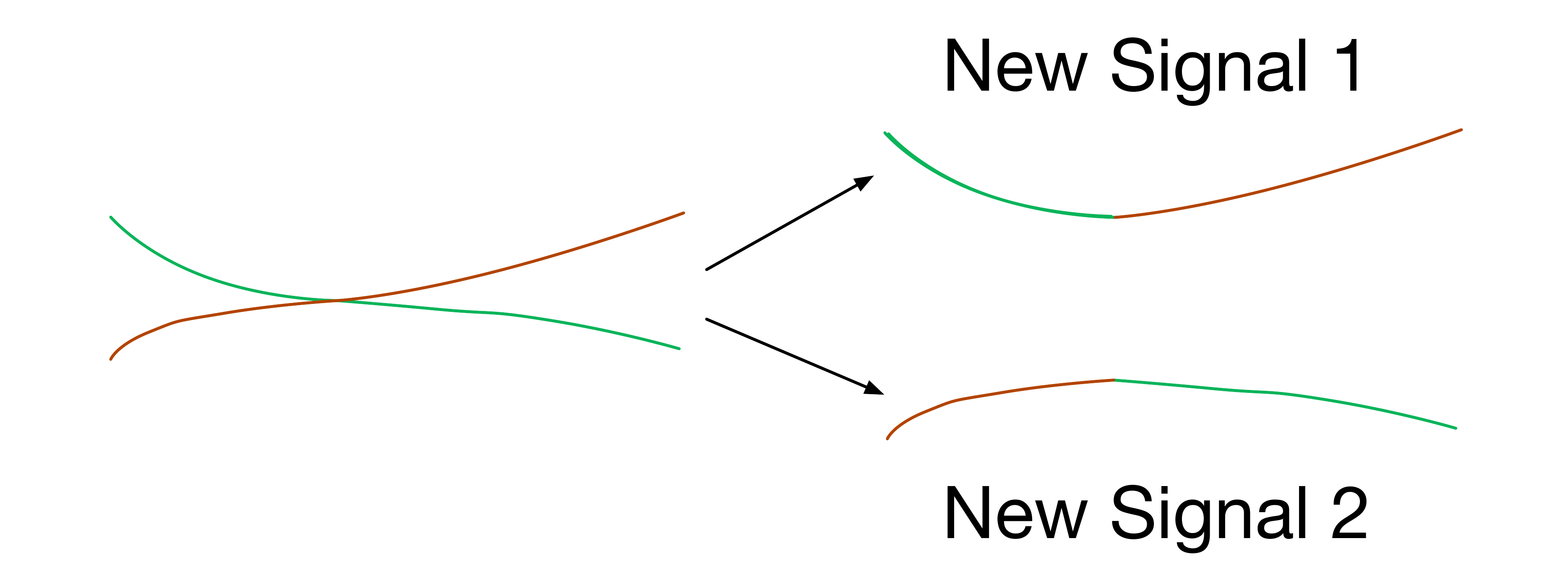}}\\
\subfloat[Sampling uniformly from the band]{\includegraphics[width=\textwidth]{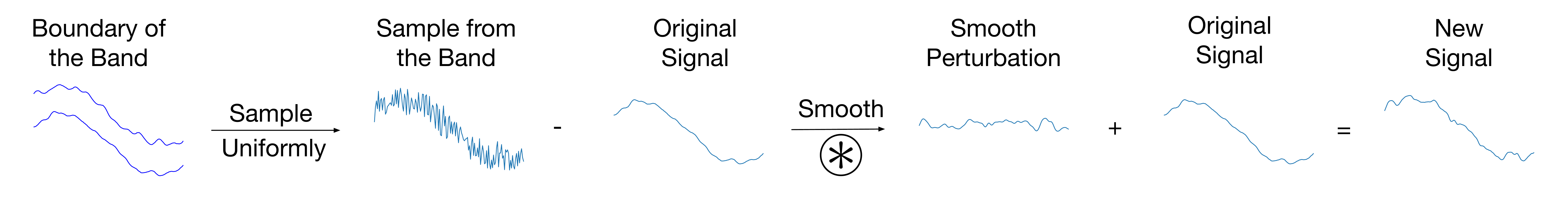}}
\caption{\label{fig:demo_ls}Demonstration of three procedures to show the existence of the adversarial examples. (a) We add small amount of Gaussian noise to the adversarial example and smooth it to create a new signal. (b) For intersected signals, we concatenate the left half of one signal with right half of the other to create a new one. (c) We sample uniformly from the band and smooth it to create a new signal.}

\end{figure}

The 1000 adversarial examples form a band. To emphasize that all the smooth signals in the band are still adversarial examples, we sample uniformly from the band to create new examples. Denote $\text{max}[t]$ and $\text{min}[t]$ to be the maximum value and minimum value of 1000 samples at time step $t$. To sample a smooth signal from the band, we first sample a uniform random variable $a[t]\sim \mathcal{U}(\text{min}[t],\text{max}[t])$ for each time step $t$ and then we smooth the perturbation. The example generated by uniform sampling and smoothing, this time is 
\[
x_{\text{adv}}'= x + Clip_{0,\epsilon}\left\{\frac{1}{m}\sum_{i=1}^m (a-x)\circledast K(s[i], \sigma[i])\right\}.
\]
We repeat this procedure 1000 times, and all the newly generated examples still cause the network to make the wrong diagnosis. We visualize the three procedures to show the existence of adversarial examples above in \Cref{fig:demo_ls}.

\clearpage
\section*{Extended Data}

\begin{figure}[h]
    \centering
    \includegraphics[width=\textwidth]{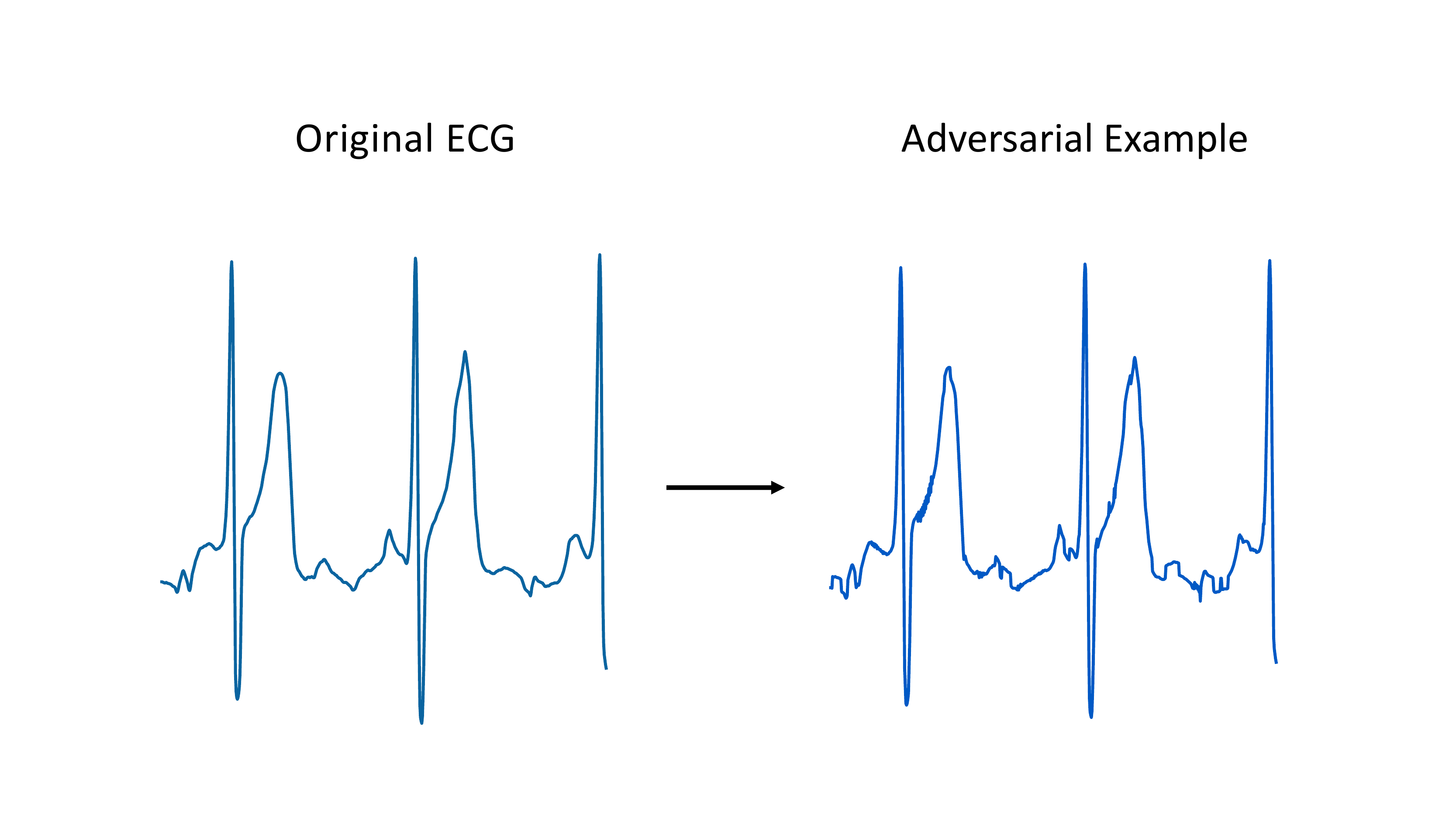}
    \caption{An adversarial example created by \glsreset{PGD}\gls{PGD} method. This adversarial example contains square waves and is not smooth. A physician reading this tracing will likely detect that this adversarial example is not a real \ECG.}
    \label{fig:pgd}
\end{figure}

\end{document}